\documentclass[twocolumn,twoside,slac_two]{revtex4}
\usepackage{graphicx}
\usepackage{fancyhdr}
\begin{document}

\title{CKM Physics from Lattice QCD}

%

\author{Paul Mackenzie}
\affiliation{Fermilab, P.O. Box 500, Batavia, IL,   USA}

\begin{abstract}
I discuss the lattice calculations relevant to recent advances
in CKM phenomenology,
focusing on those relevant to new experimental results reported
at this workshop.

\end{abstract}

\maketitle

\thispagestyle{fancy}


\section{Introduction}

Lattice QCD calculations make possible some of the most important results
in CKM phenomenology.
For example, with lattice QCD, new constraints on the CKM matrix can be
obtained from the two most important new experimental results reported at this
conference: the observation of the purely leptonic decay of the 
$B$ meson by Belle \cite{Ikado:2006fd,Ikado:2006un},
and the measurement of $B_s \overline{B_s}$ mixing by 
CDF \cite{Gomez-Ceballos:2006gj} and D0 \cite{Buchholz:2006rt}.
These quantities are examples of golden quantities for lattice QCD,
processes with stable particles and a maximum of one hadron in the lattice
box at a time.  
Other such quantities relevant to CKM phenomenology include
other meson leptonic and semileptonic decays and $M\overline{M}$ mixings.
Good prototype calculations exist in the quenched approximation (ignoring light
quark loops), and many unquenched. 
We know of no impediment to making them more and more accurate as methods
and computers continue to improve.

There are several different methods for putting quarks on a lattice
that have widely differing virtues and defects.
Wilson and clover (or improved Wilson) fermions have been the most widely
used. 
They employ a strong breaking of chiral symmetry at the lattice spacing
scale for technical reasons (to solve the fermion doubling problem).
This has lead to difficulties in reducing the light quark masses to close to
their physical values.
Staggered fermions have a chiral symmetry.
This makes possible closer approach to the physical, light quark mass limit,
smaller uncertainties from chiral extrapolation and statistics
than is possible with other methods.  However,
they also have residual fermion doubling, which in unquenched calculations
is cured by taking a root of the fermion determinant.
This procedure has been shown to make sense in every calculational test to 
it has been put, but more work is on-going to investigate it.
Two new related methods, domain wall fermions and overlap fermions,
suffer from neither of these defects, but are computationally much more 
expensive.  Practical calculations with these last two
 methods are in their infancy.
There are therefore multiple opinions among lattice theorists
about the optimal fermion method for phenomenological calculations at present.
 Several methods may have to be pushed to much higher accuracy
to test that all methods give the same results and resolve the question.

In this talk,
I will discuss mainly lattice results of direct interest to this workshop.
For recent more complete reviews, see 
\cite{Okamoto:2005zg, Hashimoto:2004hn,  Lellouch:2002nj}.

\section{Meson Decay Constants}

\begin{figure}[t]
\centering
\includegraphics[width=80mm]{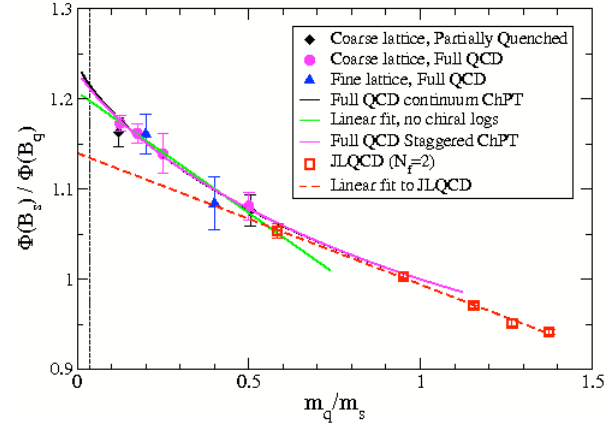}
\caption{The light quark mass dependence of 
$f_{B_s}\sqrt{M_{B_s}}/(f_B\sqrt{M_B})$
with improved staggered fermions (HPQCD, $n_f=2+1$, black symbols)
and clover fermions (JOQCD, $n_f=2$, red symbols).
Black and red lines are extrapolations using the two sets of data.
The vertical dashed line is the physical light quark mass.
\cite{Okamoto:2005zg}
} \label{fB}
\end{figure}

The practical trade-offs between Wilson and staggered fermions are illustrated
in Fig.~\ref{fB} \cite{Okamoto:2005zg}.  
It shows $f_{B_s}\sqrt{M_{B_s}}/(f_B\sqrt{M_B})$ as a function
of light quark mass.  
The vertical dashed line shows the physical quark mass.
The Wilson data \cite{Aoki:2003xb}
are shown  in red, and the staggered data \cite{Gray:2005ad} in blue.
The two methods agree where their data overlap, at $m_l\ge m_s/2$,
but curvature at light quark masses yields a significantly higher value
at the physical quark mass than is obtained from a naive extrapolation
of the Wilson data from the region $m_l\ge m_s/2$.
The staggered data yield the result
\begin{equation}
f_B=0.216(9)(19)(4)(6)\ {\rm GeV}.
\end{equation}
The uncertainties in $f_B$ and $f_{B_s}$, other than those arising from
chiral extrapolation and statistics largely, cancel out in the ratio.
HPQCD finds, with staggered fermions \cite{Gray:2005ad} in blue.,
\begin{equation}
f_{B_s}/f_B=1.20(3)(1),
\end{equation}
where the first error is from chiral extrapolation and statistics and
the second is from everything else.
This can be compared with the result that JLQCD found earlier, 
chirally extrapolating from a much larger quark mass:
$ f_{B_s}/f_B=1.13(3)({+13\atop -2} )$. \cite{Aoki:2003xb}

For $f_{B_s}$, HPQCD quotes
\begin{equation}
f_{B_s}=0.260(7)(26)(9)\ {\rm GeV}.
\end{equation}
The uncertainties in  $ f_{B_s}/f_B$ and $f_{B_s}$ are
relatively independent of each other, since  the uncertainties in the former
are dominated by statistics and chiral extrapolation, while the
uncertainties in the latter are dominated by everything else. 
$ f_{B_s}/f_B$ is more independent of $f_{B_s}$ than of $f_{B}$.

$D$ and $D_s$ meson decay constants are also of great interest because of
the improvements to the experimental numbers coming from CLEO-c.
The issues for these calculations are very similar to those for the $B$ mesons.
With staggered fermions and $2+1$ unquenched 
light quark flavors (that is, degenerate up and down quarks, and
a nondegenerate strange), Fermilab/MILC obtain 
\cite{Aubin:2005ar}
\begin{eqnarray}
f_D	&=& 201(03)_{\rm stat}(17)_{\rm sys}\ {\rm MeV}	\\
f_{D_s}	&=& 249(03)_{\rm stat}(16)_{\rm sys}\ {\rm MeV}.	
\end{eqnarray}
With clover fermions and 2 light flavors, CP-PACS obtains \cite{Okamoto:2005zg}
\begin{eqnarray}
f_D	&=& 202(12)_{\rm stat}({+20 \atop -25})_{\rm sys}\ {\rm MeV}	\\
f_{D_s}	&=& 238(11)_{\rm stat}({+07\atop -27})_{\rm sys}\ {\rm MeV}.	
\end{eqnarray}
This may be compared with the recent result from CLEO-c \cite{Artuso:2005ym}
\begin{equation}
f_D=223(17)(3)\ {\rm MeV},
\end{equation}
 assuming the canonical value of $V_{ud}$.
Alternatively, the comparison could be used to obtain an independent 
determination of $V_{ud}$.

\section{Semileptonic Decays}

\begin{figure}[h]
\centering
\includegraphics[width=80mm]{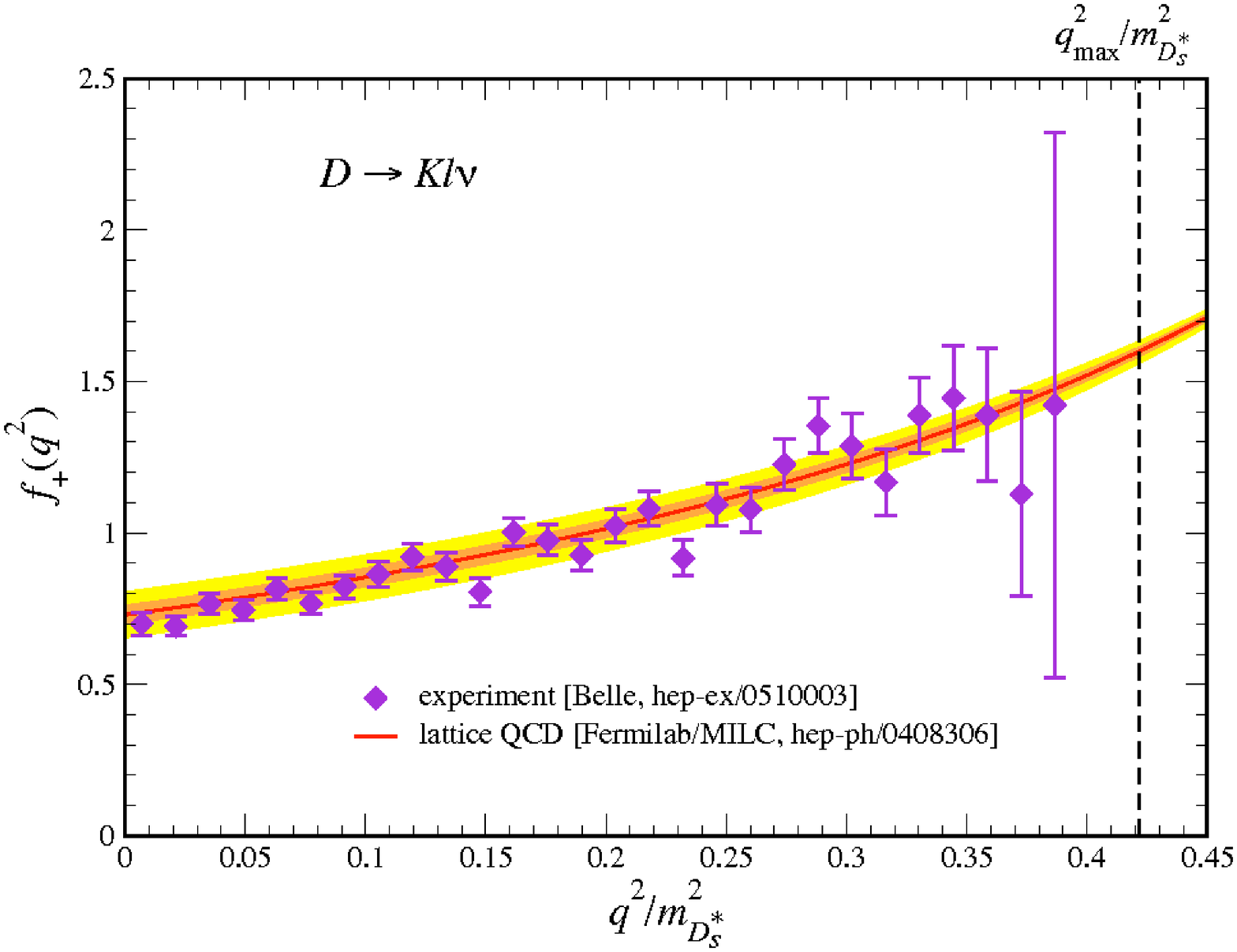}
\caption{The predicted shape of the form factor for $D\rightarrow K l \nu$
has been tested with increasing accuracy by experiment,
most recently by Belle \cite{Abe:2005sh}.
} \label{DK}
\end{figure}

The shape of the form factor for the decay $D\rightarrow K l \nu$
was obtained in 2004 by the Fermilab/MILC collaboration using
unquenched staggered fermions \cite{Aubin:2004ej}.
It has been confirmed by several experiments with increasing accuracy,
most recently by Belle \cite{Abe:2005sh} and BaBar.  (See Fig.~\ref{DK}.)
CLEO-c is expected to determine this shape with even greater precision,
yielding an even more stringent test.

The decay $B\rightarrow \pi l \nu$ is a measure of $V_{ub}$, and provides
a measurement of the height of the unitarity triangle that is competitive
with $b\rightarrow u$ inclusive decays.
For $B\rightarrow \pi l \nu$, the shape of the form
factors is a critical issue, even for extracting the CKM matrix element.
The reason is that
the uncertainties in the form factors are highly $q^2$ dependent for both
theory and experiment.
In lattice calculations, predictions do not yet exist in the high
recoil region.

It has long been knows that analyticity and unitarity can be used to constrain
the shape of form factors.
Arnesen {\it et al.} \cite{Arnesen:2005ez}
have recently applied these ideas to $B$ decay.
They show that when the variable $q^2$ is mapped into a well chosen new
variable, $z$, unitarity constrains the coefficients in the $z$ expansion
of the form factors so that only five or six terms suffice to describe the
form factors to 1\% accuracy.  They propose to to use various theoretical
methods to determine the coefficients, including lattice QCD.

\begin{figure}[h]
\centering
\includegraphics[width=80mm]{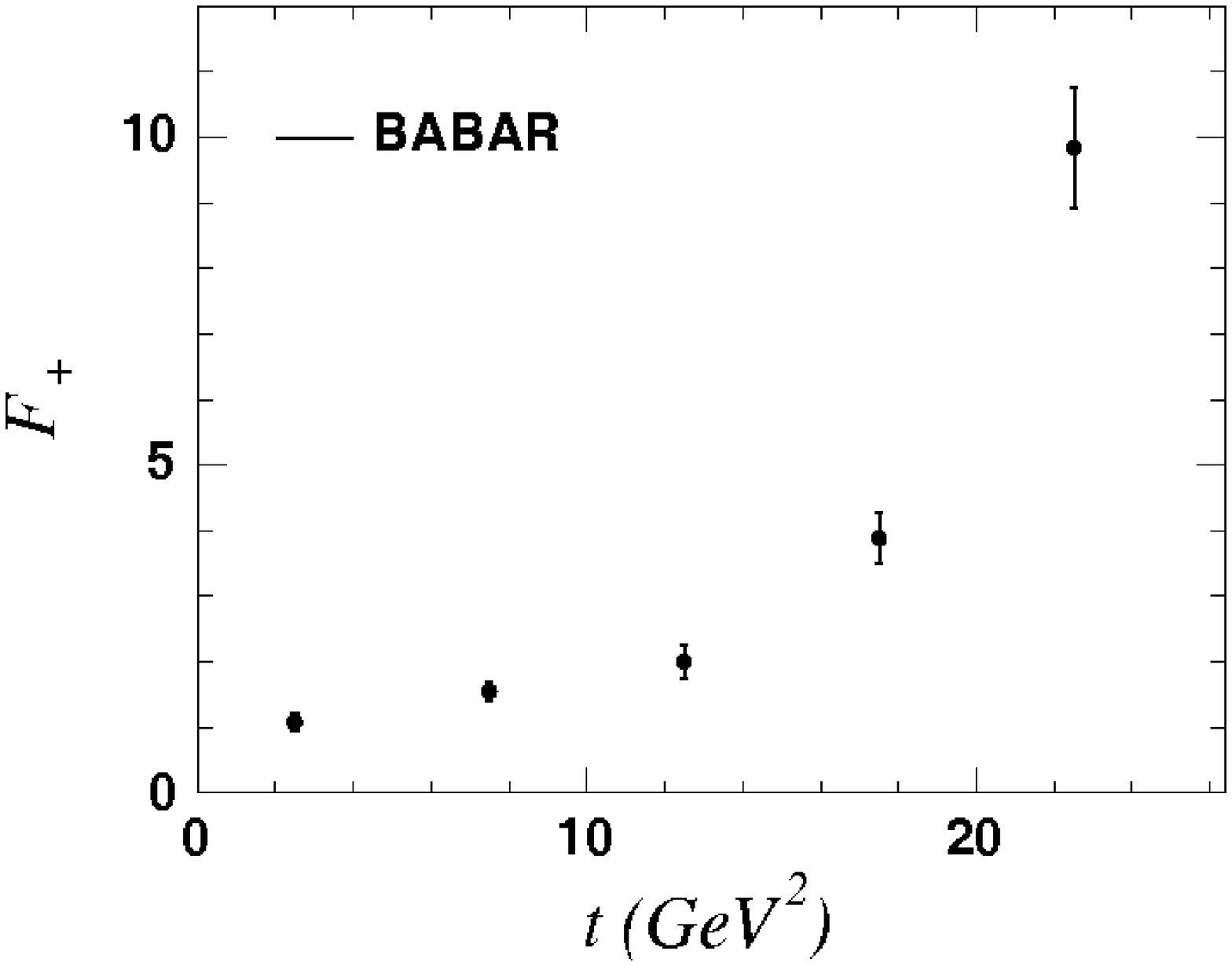}
\includegraphics[width=80mm]{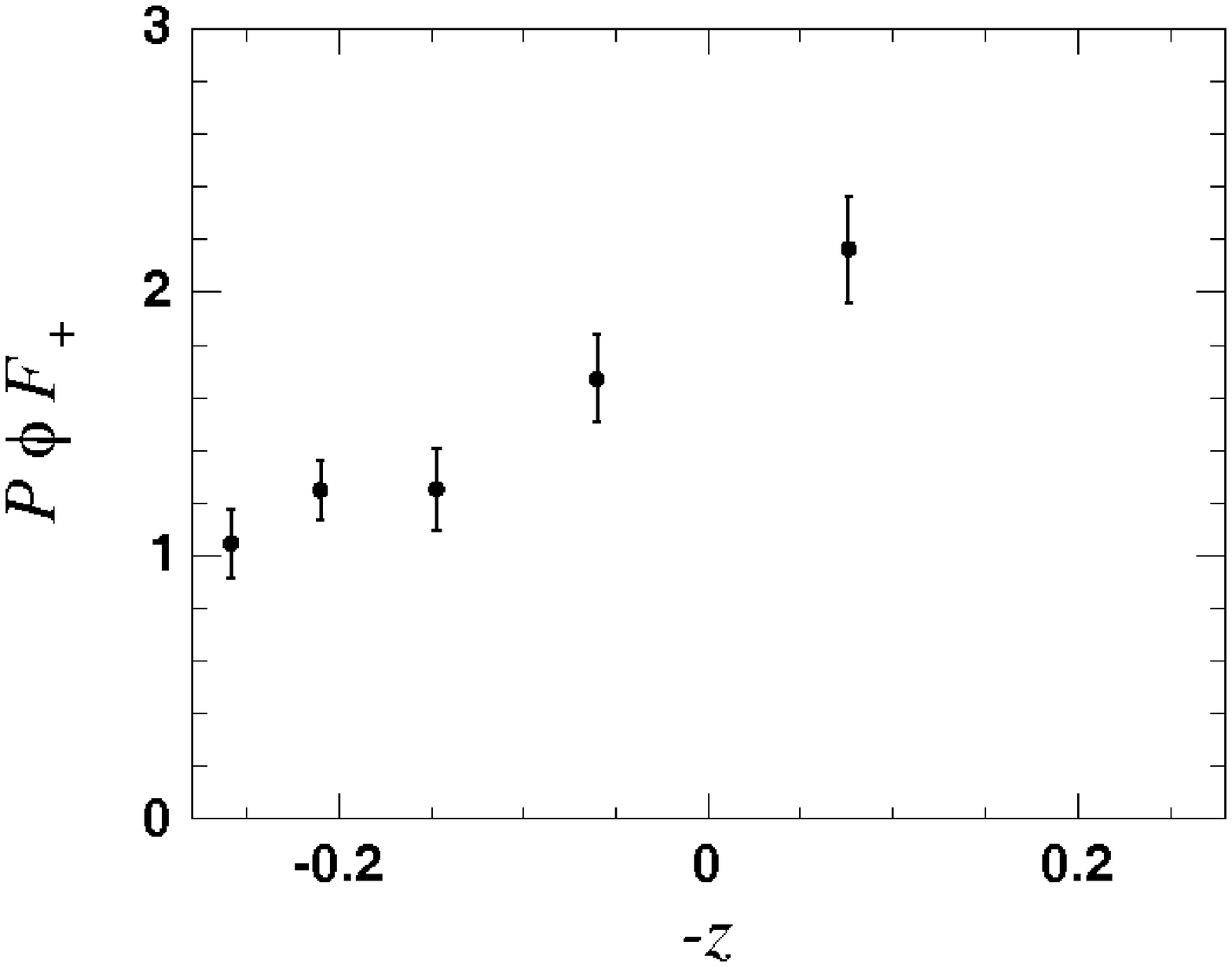}
\caption{Plotting the form facto shape for 
$B\rightarrow \pi l \nu$ decay versus a well chosen new variable $z$
(instead of $t\equiv q^2$)
produces a function that is linear to the accuracy of existing data,
and can thus be describe with only two parameters.
} \label{q2z}
\end{figure}

Becher and Hill have recently pointed out that heavy quark theory leads
to the expectation that in the heavy quark limit, the coefficients
fall as $(\Lambda_{\rm QCD}/M_B)^{1.5}$, an
 even tighter constraints on the form factors. 
\cite{Becher:2005bg}
They argue that this leads to expectation that only two or three terms
are required to describe the data to high accuracy, and show that the
BaBar data so far conform to this expectaion.  (See Fig.~\ref{q2z}.)

This expectation is also confirmed in lattice data reanalyzed the
same way  \cite{fm:2006}.
Therefore, in this expansion,
to the accuracy of the present data, comparing theory and experiment
means testing theory against experiment with the slope,
comparing the overall normalizations to obtain $V_{ub}$,
and searching for evidence of the curvature term.

\section{Meson Mixings}

Matrix elements for $B \overline{B}$ and $B_s \overline{B}_s$
mixing are conventionally parameterized in terms of the
meson decay constants and bag parameters, as in the kaon system.
What is actually calculated on the lattice, however, are the combinations
$f_B \sqrt{B_B}$ and $f_{B_s} \sqrt{B_{B_s}}$. 
Since, these combinations are what is needed for phenomenology, these are what
lattice theorists should be reporting.

Unquenched results 
 were reviewed at {\em Lattice 2005} by Okamoto  \cite{Okamoto:2005zg}.
Since unquenched results using staggered fermions had not yet appeared,
he took results for bag parameters from JLQCD using two flavors of
clover quarks,
\begin{equation}
B(m_b)=0.836(27)({\small +56\atop -62})
\end{equation}
and
\begin{equation}
\hat{B}_s/\hat{B}=1.017(16)({+56 \atop -17}).
\end{equation}
For his best estimate of the combined matrix elements, he combined
these with the HPQCD, staggered fermion results for the decay constants
already discussed, and obtained
\begin{equation}
f_B \sqrt{\hat{B_B}}=244(26)\ {\rm MeV}
\end{equation}
and
\begin{equation}
\xi=f_{B_s}/f_B \sqrt{\hat{B}_s/\hat{B}  } = 1.210({+47\atop -35 }).
\end{equation}
Following this same procedure,
one can obtain for the $B_s$
\begin{equation}
f_{B_s} \sqrt{\hat{B_{B_s}}}=294(33)\ {\rm MeV}.
\end{equation}
This quantity and $\xi$ are the combination of $B\overline{B}$ 
mixing quantities
with uncertainties that are most independent of each other.

Lifetime differences require a different operator than the one for
 the mass difference.
Unquenched calculations of matrix elements for this operator do not yet
exist.
The required operator has been given as part of the complete 
set of four-quark operators, including those arising in 
supersymmetric theories, and initial results have been reported in
the quenched approximation \cite{Becirevic:2001xt}.

\section{Effect on the $\rho-\eta$ plane}

\subsection{Effects of Lattice Fermion Methods}

In this talk, I have emphasized staggered fermion results
because they are for the most part fully unquenched, with the right number
of light quark flavors, and with light quarks masses much closer to their
physical values.
Other lattice theorists have a preference for other fermion methods,
and the lattice community is not yet settled on the optimum approach.
Another widely used set of numbers for $B\overline{B}$ mixing parameters
comes from the CERN CKM study of 2003 \cite{Battaglia:2003} 
(based on the 2002 review of Lellouch \cite{Lellouch:2002nj})
\begin{eqnarray}
f_B \sqrt{\hat{B}_B}	&=&  235(33)({0\atop 24})	\\
f_{B_s} \sqrt{\hat{B}_{B_s}}	&=&  276(38)	\\
\xi	&=& 1.18(4)({12\atop 0}).
\end{eqnarray}
These were based on a set of mostly clover lattice results, 
most in the quenched approximation but with some two flavor results
used to extrapolate to the three-flavor theory.
The results are compatible with the staggered results already discussed,
but with larger uncertainties.

\begin{figure*}[t]
\centering
\includegraphics[width=135mm]{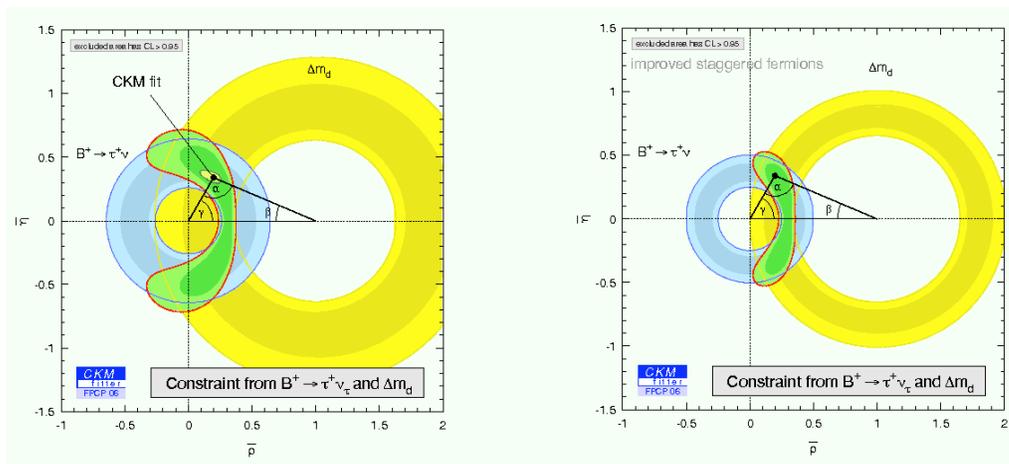}
\caption{Constraints on the $\rho-\eta$ plane derived  from 
 $B$ meson leptonic decay 
and $B \overline{B}$ mixing.
The bounds in the left figure employ clover fermions, those on the right
employ staggered.
} \label{charles}
\end{figure*}

The effects of the smaller uncertainties of staggered fermions were shown
by J. Charles at this workshop \cite{Charles:2006}.  Fig.~\ref{charles}
shows the bounds on the $\rho-\eta$ plane derived solely from 
 $B$ meson leptonic decay 
and $B \overline{B}$ mixing.
The bounds in the left figure employ the clover fermion results
just discussed, those on the right employ the staggered fermion results
discussed earlier.

\subsection{CKMfitter and UTFit}
There are two groups producing widely cited global fits
to CKM data: CKMfitter \cite{ckmfitter} and UTfit \cite{utfit}.
The groups employ different statistical methods, and obtain somewhat
different results.
For example, for $\Delta m_s$ predicted without incorporating the 
experimental $B_s \overline{B}_s$ mixing results, UTfit obtains
21.5(2.6) ps$^{-1}$, while CKMfitter reports $21.7({5.9\atop 4.2})$ ps$^{-1}$.
This is puzzling, since
one would expect this quantity to be sensitive mainly to lattice uncertainties,
and both groups take the lattice results of Ref.~\cite{Battaglia:2003}
as the starting point.
One possible difference is that UTfit lists $f_{B_s} \sqrt{\hat{B}_{B_s}}$ and 
$\xi$ as inputs, while CKMfitter lists $f_{B} \sqrt{\hat{B}_{B}}$ and $\xi$.
The latter combination contains highly correlated uncertainties that must
be treated with care.
However,
a more significant difference seems to arise from 
statistical methods and treatments of combinations of uncertainties,
rather than differences of lattice inputs.
These differences remain yet to be resolved.

\section{Outlook}

Lattice calculations are playing an essential role in enabling
some of the most important results in the CKM experimental program.
Most of the key lattice calculations for CKM physics involve
single-hadron processes with hadronically stable meson, which are among the
most solid current lattice calculations.
Significant issues are currently being worked through,
for example, the best fermion method for phenomenological calculations,
the best way to incorporate lattice results into Standard Model
global fits,
and the best way compare theory and experiment in semileptonic decays.
Lattice  methods are currently in a state of productive ferment,
with several different methods for unquenched lattice fermions under active
investigation by various groups.
Lots of progress is being made in algorithms for these various methods
\cite{Luscher:2005mv}.
The computing power being applied to lattice phenomenology is rising
exponentially \cite{Wettig:2005zc}.
There are excellent prospects for further progress in lattice CKM
phenomenology.


\bigskip 

\end{document}